# Algorithm for Sector Spectra Calculation from Images Registered by the Spectrometer Airglow Temperature Imager

*Atanas Marinov Atanassov*

*Solar-Terrestrial Influences Institute, Bulgarian Academy of Sciences, Stara Zagora Department, Bulgaria,*
At_M_Atanassov@yahoo.com

*In memory of our late colleague Krasimir Kanev*

The Spectral Airglow Temperature Imager is an instrument, specially designed for investigation of the wave processes in the Mesosphere-Lower Thermosphere. In order to determine the kinematic parameters of a wave, the values of a physical quantity in different space points and their changes in the time should be known. As a result of the possibilities of the SATI instrument for space scanning, different parts of the images (sectors of spectrograms) correspond to the respective mesopause areas (where the radiation is generated).

Algorithms for sector spectra calculation are proposed. In contrast to the original algorithms where twelve sectors with angles of 30° are only determined now sectors with arbitrary orientation and angles are calculated. An algorithm is presented for sector calculation based on pixel division into sub pixels. A comparative results are shown.

## Introduction

The study of the dynamic processes in the Mesosphere/Low Thermosphere is significant for understanding the atmosphere energetics on a global scale. The investigation of the mechanical wave processes in the atmosphere with optic instruments is possible indirectly by determining the periodic variations in some scalar 2-dimensional fields (temperature and emission intensity). Originally, these investigations were conducted by spatial scanning ground-based photometers [1] registering the OH infrared emissions. The optic axis of these instruments is sequentially directed to different points of the nocturnal sky where the registered signal is coming from.

Recently, the Spectral Airglow Temperature Imager (SATI) has been developed for internal gravity waves investigation [2]. In contrast to the spatial scanning photometers, employing the classical objective, the optic scheme of SATI includes a special conic mirror. The instrument accepts the light from the ring-shaped segment which is projected on the sky by this mirror. By narrow interference filters, the airglow emissions connected with $O_2(b^1\Sigma_g^+)$ and $OH(X^2\Pi,v)$ are dispersed to produce ring-shaped spectral patterns. The maxima of these airglow emissions are at 94 and 87 km, respectively, which is the altitude of the mesopause.

For determination of the temperature and emission intensity, the images are divided into sectors which correspond to parts of the ring-shaped sky segment where the light comes from [3, 4].

On the basis of the registered airglow spectra for different mesopause areas, the temperatures and the gas emission intensities in these areas are determined. By applying the frequency analysis to the temperature and emission intensity time series from the entire nocturnal séance, the amplitudes, periods and phases of statistically reliable harmonic components are determined [3]. Finally, on the basis of the found components, which have sufficiently close periods and amplitudes, the wave parameters (magnitude and direction of horizontal velocity vector, wave longitude) are determined [3].

A version SATI-3SZ was developed at the Stara Zagora Division of the Solar-Terrestrial Influences Laboratory in collaboration with CRESS Laboratory [5].

An algorithms for data processing registered with this instrument [6, 7] are developed. This paper presents approaches for calculation of sector spectra. Some comparative results are also presented.

## Sector spectra calculation by classical algorithms for SATI image processing

In original algorithms [4] of SATI data processing the images (Fig. 1a) are divided on 12 sectors with equal angle by 30° each (Fig.1b). This is possible after determination of the image centre coordinates $(i_0, j_0)$. These coordinates are calculated with precision of a whole pixel along each of the two directions of the axes Ox and Oy. Every pixel at a distance less than the image radius (~128ps) is checked in which sector falls according to the angle between its radius vector and the basic direction of the axis Ox

$$\gamma = a\tan((i-i_0)/(j-j_0)),$$

Besides, according to the distance to the image centre p (in pixels), the values of the registered pixel intensity is summed up with the intensities of all pixels located at the same distance:

if $\gamma \in (\gamma_{1,k}, \gamma_{2,k})$, then

$$S_k(p) = S_k(p) + I(i,j), \quad r = \sqrt{(i-i_0)^2 + (j-j_0)^2},$$

where $S_k$ is k-th sector sum in the pixel space and p accepts the rounded value of r. Finally, all produced sums for the sector are averaged in accordance with the number of pixels at the same distance from the centre; in the same way the value of a spectrum element (Fig. 1c) is determined for the respective sector

$$S_p^k = S_p'^k / n_p^k, \quad n_p^k = \sum_{p=\sqrt{(i-i0)^2+(j-j0)^2}} I_{i,j}$$

The spectra measured by the SATI instrument are not linear in relation to the wavelength [7]. The spectra are presented as series of intensities as a function of the distance to the image centre. The correspondence between the serial number of each pixel and the wavelength is significant for the interpretation of the registered spectra, not for their extraction from the images. Thus, each spectrum is formed as a series of values of registered intensities for the respective wavelength, corresponding to the respective pixel.

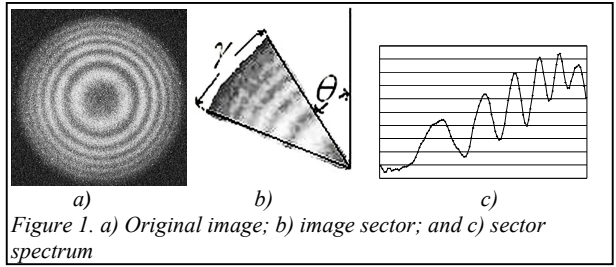

*Figure 1. a) Original image; b) image sector; and c) sector spectrum*

Each sector spectrum is compared with preliminary calculated spectra (synthetic spectra) for different temperatures of the radiating gas (in our case rotation spectra of $O_2$), convoluted with the transmittance function of the employed interference filter.

This approach allows change of the sector angle in a close connection with the number of sectors N: $\gamma^0 = 360°/N$.

Here we will note only that each of the calculated sector spectrum is smoothed for noise removal in radial direction. In this way practically we have a space image filtration. This filtering is made with five points meaning.

**A new approach proposed for sector spectra determination**

According to the proposed approach for sector spectra calculation, the sector angle $\gamma^n$ may be chosen in an arbitrary manner within certain boundaries (even wider than $\gamma^0$) and according to the statistical error. Since efficiency is required in the pixel selection process and for the check of the pixels in a given sector, constraining radii will be used whose canonical equations are the following:

$$\begin{Vmatrix} y = k_1 x, & k_1 = tg\theta \\ y = k_2 x, & k_2 = tg(\theta + \gamma) \end{Vmatrix}$$

In contrast to the original approach for image centre coordinate determination, an algorithm was developed for its calculation with a precision higher than one pixel [7]. That is why the coefficients d1 and d2 in the couple of equations (1) are real numbers.

In order to choose the pixels restricted by the two segment radii, their indices are changed, depending on the following versions (Fig. 2):

**Version I:** the sector is entirely in one of quadrants I, II, III or IV only and we can write down for index by Ox axis

$i \in (NINT(\sin g(1, \cos \varphi_1)), NINT(R * MAX(\cos \varphi_1, \cos \varphi_2)))$, for $\cos \varphi_1 > 0$

$i \in (NINT(\sin g(1, \cos \varphi_1)), NINT(R * MIN(\cos \varphi_1, \cos \varphi_2)))$, for $\cos \varphi_1 < 0$

**Version II:** the sector falls into I and IV or II and III quadrants simultaneously. Then for the changing of the index by Ox axis

$i \in (NINT(sign(1, \cos \varphi_1)), NINT(sign(R, \cos \varphi_2)))$.

We will note only that in the versions described here, $sign(\cos(\varphi_1)) = sign(\cos(\varphi_2))$. The function NINT() is intrinsic function in the Fortran programming language which returns the nearest integer to the argument [8]. The functions MAX() and MIN() return the maximum or minimum value respectively of the arguments.

For the change of index by Oy axis in the above two versions we can write down

$j \in (NINT(k_1.i), NINT(k_2.i)), k_1.i < k_2.i$

or

$j \in (NINT(k_2.i), NINT(k_1.i)), k_2.i < k_1.i$

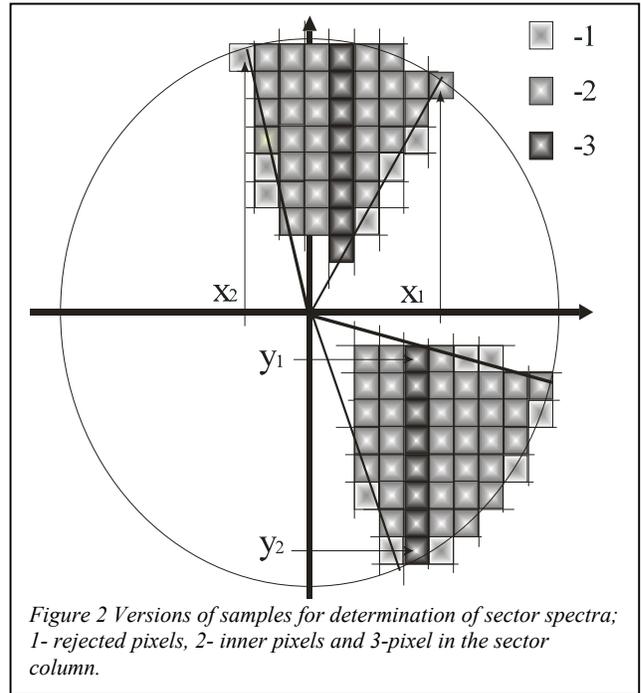

*Figure 2 Versions of samples for determination of sector spectra; 1- rejected pixels, 2- inner pixels and 3-pixel in the sector column.*

**Version III:** the sector falls into I and II or III and IV quadrants simultaneously. Then

$i \in (NINT(R.\cos\varphi_1), NINT(R.\cos\varphi_2)), \cos\varphi_1 < \cos\varphi_2$

or

$i \in (NINT(R.\cos\varphi_2), NINT(R.\cos\varphi_1)), \cos\varphi_1 > \cos\varphi_2$

For the index change along the Oy axis in the last third version we have for $\cos\varphi_1 < \cos\varphi_2$

за $i \in (NINT(R.\cos\varphi_1), 0) \quad j \in (k_1.i, \sqrt{R^2 - i^2})$

и за $i \in [0, NINT(R.\cos\varphi_2) \quad j \in (k_2.i, \sqrt{R^2 - i^2})$

Analogously, the boundaries in which the indices change and by $\cos\varphi_1 > \cos\varphi_2$ can be determined.

The distance of every pixel to the image centre, as above, determines the process of summing and averaging

$$S_p = \frac{1}{M_p} \sum I_{i,j}, p = NINT(\sqrt{i^2 + j^2})$$

where $S_p$ is the $p^{th}$ element of the sector spectrum, $M_p$ is the number of all pixels at a distance of p pixels from the image centre.

Actually, this approach for averaged sector spectra calculation like the original one with twelve sectors only, represents one-dimensional filtering with one of the most simple and widely used filter [9], however, applied on a series of elements (pixels) disposed on curved lines- circles. What is special in our case is the formation of one-dimensional (curvilinear by circle) pattern on two-dimensional image for simultaneous calculation of the whole spectrum.

In contrast to the original algorithm [4,10] where the sector spectra produced after averaging are additionally smoothed with a mean filter by five points, a double pass filtering was applied by three points to reduce the influence of the long window on values around the minima and maxima [9, 6].

**Some results from the application of the algorithm**

There are differences around the crests and valleys when comparing the spectra with various sector angles (Fig. 3). Every value of an element of the sector spectre $S_p$ is determined by averaging the pixels disposed on a circle arc with radius **p** pixels. Basically, this approach is equivalent to the application of the so-called "moving average".

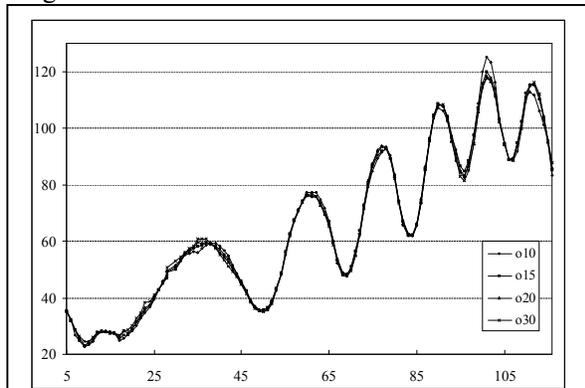

*Figure 3. Sector spectra, obtained at different sector angles: o10 -10°, o15 -15°, o20 -20° and o30 -30°.*

It is known that this kind of averaging the adjacent values in one image plays the role of a high-frequency filter and cleans the noise. It is also known that this kind of filtering influences weakly the low frequency components. By a 1- or 2-dimensional function in the vicinities around the local extrema, along with the noise, the values of the minima and maxima also change. The longer the filter window (the sector angle in our case), the better the cleaning. However, the probability for pixels with abnormal values to get into the sector increases.

**Sector specters determination by pixels fragmentation**

A version of the presented algorithm has been developed which treats every pixel as a matrix from (mxm) sub-pixels and the summing of values $\frac{I_{i,j}}{m^2}$ is accomplished according to the distance to the image centre. The exact coordinates of the image centre and the distance of every sub-pixel to it are important for the calculation of the sector spectrum. Fig. 4 shows two sector spectra with equal determining angles which have been calculated by the two approaches.

It is obvious that there is a shift between the two spectra which is within the frame of two pixels. This shift is not uniform on the whole image and depends on the fractions of the determined coordinates of the image centre.

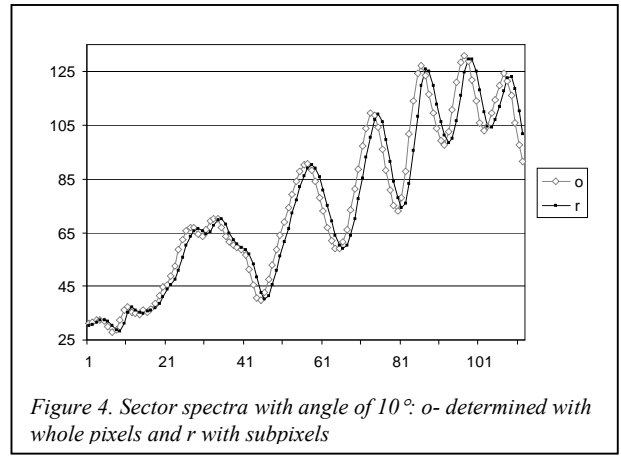

*Figure 4. Sector spectra with angle of 10°: o- determined with whole pixels and r with subpixels*

**Conclusion and future work**

The presented approach for sector spectra determination is characterized with flexibility in relation to the possibilities for free selection of the sector angle and the number of sectors. With this approach it would be possible to process only part of the entire image or to apply a different standard to a selected part of it.

The version of the algorithm for sector spectra calculation based on pixel value fragmentation into sub pixels allows to use the more precise image centre determination within the frames of using one pixel. Analysis is envisaged to establish whether the more precise calculation of the sector spectra would lead to improvement of the sector temperature determination.

The increase of the sector angle leads to more calculations when determining the sector spectra. When angle $\gamma > 360°/N$ the sectors are overlapped. A recurrent algorithm is being developed which will be employed to determine every next sector spectrum on the basis of the previous one. This is analogous with the fast moving average algorithm [9] whose width is known and which represents an effective approach.

When determining sector spectra by averaging the pixel values which are at equal distances from the centre within the frame of a sector angle, a very popular filtering method is applied, known as a moving average mean, rectangular, box, uniform filter. It would be interesting to investigate the possibilities to construct other algorithms for sector spectra determination on the basis of other approaches for filtering (median filtering, Savitzki-Golay smoothing [11]) as well as their role for the precision of the determined temperatures.

**Acknowledgements:** The author would like to thank Dr. M.G. Shepherd for the support and encouragement and Mrs. Kr. Takucheva for the technical assistance.